\documentclass[prc,showpacs,showkeys,superscriptaddress,nofootinbib,twocolumn,floatfix]{revtex4}
\usepackage{graphicx,color,mathrsfs,amsmath,amssymb,amsthm,amsopn,bm}
\usepackage[sort&compress]{natbib}
\newlength{\lslashl}

\def\w2{\tilde w^2}
\def\ws2{1}

\def\unitmatrix{\hbox{$1\hskip -1.2pt\vrule depth 0pt height 1.6ex width 0.7pt
                  \vrule depth 0pt height 0.3pt width 0.12em$}}
\newcommand{\ul}{(u\cdot l)}
\newcommand{\uq}{(u\cdot w)}
\newcommand{\ql}{(l\cdot w)}
\usepackage[normalem]{ulem} 

\renewcommand\sout{\bgroup \color{red} \ULdepth=-.5ex \ULset}

\begin{document}

\title{Selfconsistent descriptions of vector-mesons in hot matter revisited}

\author{F. Riek}\thanks{e-mail: friek@comp.tamu.edu}
\affiliation{Cyclotron Institute and Physics Department, Texas A\&M
  University, College Station, Texas 77843-3366, USA} \author{J\"o{}rn
  Knoll}\thanks{e-mail:j.knoll@gsi.de} \affiliation{GSI
  Helmholzzentrum f\"ur Schwerionenforschung GmbH, Planckstr. 1, 64291
  Darmstadt, Germany}

\begin{abstract}
  Technical  concepts are presented  that improve  the selfconsistent
  treatment  of  vector-mesons  in  a  hot  and  dense  medium.  First
  applications concern an interacting  gas of pions and $\rho$ mesons.
  As  an extension  of  earlier studies  we  thereby include  RPA-type
  vertex corrections and further  use dispersion relations in order to
  calculate the real part of the vector-meson selfenergy. An improved
  projection   method  preserves  the   four  transversality   of  the
  vector-meson  polarisation  tensor  throughout  the  selfconsistent
  calculations,  thereby  keeping   the  scheme  void  of  kinematical
  singularities.
\end{abstract}

\date{\today}
\maketitle

%
\section{Introduction}
%

The  study  of  the  in-medium  properties  of  hadrons  has  received
considerable attention during  the recent years.  From chiral-symmetry
considerations one expects strong changes in the spectral distribution
of particles  when approaching the phase transition  form the hadronic
phase into  the quark-gluon plasma.  Apart  from interesting many-body
effects like particle-hole  excitations, scattering off particles from
a heat bath or Landau-damping  which are present at finite temperature
and  density  such investigations  also  allow  to  gain insight  into
fundamental  aspects  of   quantum  chromodynamics  (QCD),  cf.   e.g.
\cite{Leupold:2009kz,Rapp:2009yu,Tserruya:2009zt}.  A special research
focus  has been  on  vector-mesons studied  through  their decay  into
electron--positron or muon--antimuon pairs, called dileptons.  Void of
complicated  final-state   interaction  effects  such  electromagnetic
signals directly observe the center of the reaction zone and therefore
allow to  get an unperturbed view  on the medium  modifications of the
vector mesons.  Here  several collaborations studied vector-mesons and
especially  the  $\rho$-meson  in nucleus-nucleus  \cite{Eberl:2005ks,
  Matis:1994tg,    Wessels:2002ha,   Angelis:1998im,   Arnaldi:2006jq,
  Adamova:2006nu} and hadron-nucleus \cite{Naruki:2005kd,Ozawa:2000iw}
collisions.   In such  experiments  a significant  enhancement in  the
dilepton rates was  observed in the invariant pair-mass  region of 300
to  600 MeV,  compared  to estimates  from  straight extrapolation  of
elementary processes.  These observations triggered quite a variety of
explanations, which range from a  lowering of the $\rho$-meson mass to
a  significant  increase of  its  damping width  \cite{Asakawa:1992ht,
  Herrmann:1993za,    Friman:1997tc,    Rapp:1997fs,    Peters:1997va,
  Urban:1999im,     vanHees:2000bp,    Post:2000qi,    Cabrera:2000dx,
  Riek:2004kx}.   Presently  the  high  precision  data  of  the  NA60
collaboration  \cite{Arnaldi:2006jq}  are  best described  assuming  a
strong   broadening  of   the   $\rho$-meson  width   in  the   medium
\cite{vanHees:2007th,Rapp:1999ej}.  This  seems also to  be compatible
with  results  from   hadron-nucleus  collisions  and  photo-production
experiments \cite{Ozawa:2000iw,Clas:2007mga,Wood:2008ee}  where also a
broadening  is favored and which could be  explained by  theoretical models
\cite{Effenberger:1999ay,Muhlich:2002tu,Wood:2008ee,Riek:2008ct,Riek:2010gz}.    On
the other hand the  experimental results of Ref.  \cite{Naruki:2005kd}
favor a mass shift of the $\rho$-meson with only little broadening.

So  far most theoretical  investigations were  done on  a perturbative
level   \cite{Urban:1999im,  Rapp:1999qu,   Rapp:1999us,  Rapp:1999ej,
  Post:2003hu, Post:2000qi,  Cabrera:2000dx}.  This allows  to include
large  numbers of  excitation modes  contributing to  the $\rho$-meson
spectral         function.          selfconsistent         treatments
\cite{vanHees:2000bp,Riek:2004kx,Ruppert:2004yg,Riek:2006vq,RR-Eratum}
showed interesting new effects. However  so far the model space in the
latter  studies   was  rather   limited  and  mesonic   systems  where
investigated only.   In a previous work  \cite{Riek:2004kx} we already
improved  the situation by  considering baryon  effects on  the pionic
modes  in the  medium.   Significant progress  in  the description  of
mesons and  baryons on the vacuum  level has been  achieved by coupled
channel approaches \cite{Penner:2002ma,Penner:2002md,Lutz:2001mi}.  In
\cite{Lutz:2001mi} this input was  then used to draw conclusions about
the in-medium behavior of  vector-mesons. Here quite different effects
as compared  to the calculations of Post  et.  al.  \cite{Post:2000qi}
were  found due to  the smaller  coupling of  the $\rho$-meson  to the
$N^*$(1520)  resonance.  Thus  despite the  recent success  of several
models   in  explaining   the  experimental   data   more  theoretical
investigations are needed in order  to understand the dynamics in more
detail.

In this work we will  concentrate on some conceptual aspects which are
important  for the  improvement of  selfconsistent  descriptions.  The
first concerns  the treatment of vertex  corrections initially studied
already in  \cite{Asakawa:1992ht,Herrmann:1993za}.  For baryon systems
, including self-consistency, 
their     special      role     was       shown     in     Refs.
\cite{Lutz:2007bh,Korpa:2008ut,Riek:2008uw}.   Here   we  shall  study
their role in  mesonic systems with the perspective  to generalize the
techniques   to   the   coupled   system   of   mesons   and   baryons
\cite{Riek:2004kx,Rapp:1999ej}.  As a second point we will address the
issue  of renormalization.   So far  in all  selfconsistent treatments
\cite{vanHees:2000bp,    Riek:2004kx,   Ruppert:2004yg,   Riek:2006vq,
  RR-Eratum} the  real parts of the selfenergies  were neglected.  For
renormalizable   theories  it   was  shown   in  \cite{vanHees:2001ik,
  VanHees:2001pf, vanHees:2002bv} how  a proper renormalization has to
be performed.  However since we are working within the framework of an
effective field theory  such a procedure is not  applicable and we use
dispersion relations  or formfactors  instead.  For vector  mesons one
has  to face  the additional  problem that  due to  the  violation of
certain Ward-identities also longitudinal  modes will be propagated in
selfconsistent approximation  schemes.  Several methods  were proposed
\cite{vanHees:2000bp,    Riek:2004kx,   Ruppert:2004yg,   Riek:2006vq,
  RR-Eratum} in order  to cure this problem.  They all  have one or an
other conceptual drawback  \cite{Riek:2006vq} especially linked to the
appearance of  kinematical singularities.  Here  we will show  how the
scheme introduced in \cite{vanHees:2000bp} can be extended in order to
deal with this problem.

The paper  is organized as follows. In  section \ref{approximation} we
provide an overview of the  model and the approximation scheme used to
develop  our  techniques  within  the  selfconsistent  framework.   In
section  \ref{SecDetailVec} we  than  go into  more  detail about  the
calculations,  however,   deferring  more  technical  aspects  to  the
appendices.    The  results   will  than   be  presented   in  section
\ref{results} before concluding.

%
\section{The approximation scheme\label{approximation}}
%

The Lagrangian defining the interaction between the isospin triplet
fields of pions and
$\rho$-mesons, $\pi$ and $\rho$,  is given by
\cite{Urban:1998eg}
\begin{eqnarray}
&&{\mathcal L}_{\pi\rho}^{vector}
=\frac{1}{2}(\partial_{\mu}-ig\rho_{\mu}T^{1})\, \pi\cdot
(\partial_{\mu}-ig\rho_{\mu}T^{1})\,\pi \nonumber\\ 
&&\qquad\qquad\,-\frac{1}{2}m_{\pi}^{2}\pi\cdot\pi
-\frac{1}{4}\rho_{\mu\nu}\rho^{\mu\nu} 
+\frac{1}{2}m_{\rho}^2\rho_{\mu}\rho^{\mu}
\label{L-vector} 
\end{eqnarray}
in  vector  representation  (see  e.g.   \cite{Leupold:2006bp}  for  a
discussion of  tensor representation).  Here the  isospin structure of
the terms  is not  explicitly given. The  vector meson couples  to the
pions   through   the  fully   anti-symmetrized   tensor  in   isospin
$T^{1}_{abc}=-i\epsilon_{abc}$ with  isospin indices $a,b,c$.  Thereby
$\rho_{\mu\nu} =  \partial_{\mu}\rho_{\nu} - \partial_{\nu}\rho_{\mu}$
denotes  the  vector-meson  field  strength  tensor.   The  parameters
$g=5.3$  and $m_{\rho}=773$  MeV are  adjusted to  the electromagnetic
form factor of the pion and  compare quite well to the values found in
perturbative  calculations.   In  order  to avoid  contributions  from
non-physical modes such as ghosts, the vector meson propagator will be
treated  in the  unitary gauge  limit, which  pushes  all non-physical
modes  to infinite  masses.  The  expressions  for the  free pion  and
$\rho$-meson propagator then read
\begin{eqnarray}
\hspace*{-1cm}&& D^{0}(w)=\frac{1}{w^2-m_{\pi}^2+i\epsilon}\,,\quad
G^{\,0}_{\mu \nu}(w)=\frac{g_{\mu
    \nu}-\frac{w_{\mu}\,w_{\nu}}{m_{\rho}^2}}{w^2-m_{\rho}^2+i\epsilon}. 
\label{prop-vector-mesons}
\end{eqnarray} 
The selfconsistent retarded propagators  $D$ and $G_{\mu \nu}$ of pion
and $\rho$-meson  are given as solutions  of the coupled  set of
Dyson equations
\begin{eqnarray}
 G_{\mu \nu}(w,u) &=& G^{\,0}_{\mu \nu}(w)+ G^{\,0}_{\mu
   \alpha}(w)\,\Pi^{\alpha \beta}_{(\rho)}(w,u)\, G_{\beta
   \nu}(w,u)\,,\nonumber\\ 
 D(w,u)&=&D^{0}(w)+D^{0}(w)\,\Pi_{(\pi)}(w,u)\,D(w,u)\,.
\label{Dyson-pi-rho}
\end{eqnarray}
In  \cite{vanHees:2000bp,Riek:2004kx,Ruppert:2004yg}  only the  lowest
order  selfenergy diagrams given  by the  interaction (\ref{L-vector})
were included. As an extension to this approach we will now also study
a  first set  of  vertex  corrections which  proved  important in  the
description   of    baryons   \cite{Asakawa:1992ht,   Herrmann:1993za,
  Lutz:2007bh,   Korpa:2008ut,  Riek:2008uw}.    We  start   from  the
assumption that all soft modes of  the system have to be resumed while
the hard  modes can  be effectively treated  as local  point vertices.
Then the key ingredient of our calculation is the correlation loop
\begin{eqnarray}
  &\mathbf{\chi}^{\mu\nu}&=\text{\parbox{22\unitlength}
{\includegraphics[scale=0.5]{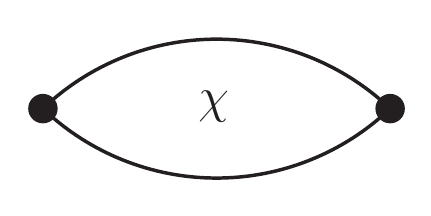}}}
=\text{\parbox{22\unitlength}{\includegraphics[scale=0.5]
{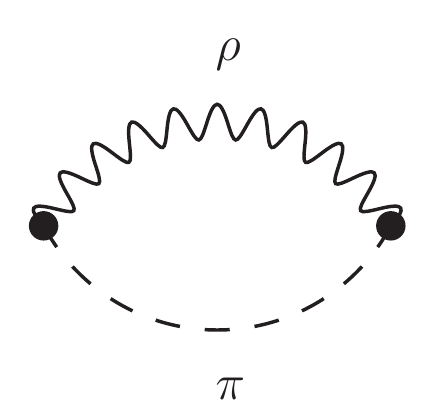}}}\,. 
\label{correlation-loop}
\end{eqnarray}
In  a  relativistic   treatment  it  takes  the  form   of  a  Lorentz
polarization tensor\footnote{At  this point also  nucleon-hole or more
  general correlation  loops could be included by  extending the matrix
  structures   along    the   lines   of    Ref.    \cite{Lutz:2007bh,
    Korpa:2008ut}. Possibly one then also has to allow for a more envolved matrix structure in the coupling $g$ which in our case is just a unit matrix.}.  From the interaction Lagrangian (\ref{L-vector})
one can then construct the following resummed correlation functions
\begin{eqnarray}
\Pi^{\mu\nu}=&\text{\parbox{25\unitlength}{
\includegraphics[scale=0.6]{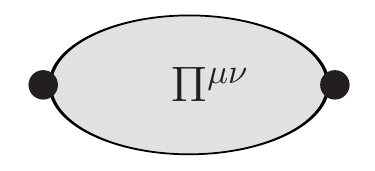}}}&
=\left[\chi\cdot\left(1-g\cdot\chi\right)^{-1}\right]^{\mu\nu}\nonumber\\
\Gamma^{\mu\nu}=&
\text{\parbox{25\unitlength}{
\includegraphics[scale=0.6]{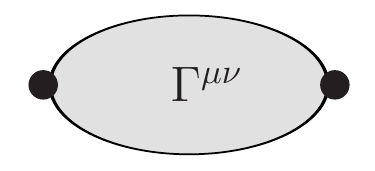}}}&
=g\cdot\Pi^{\mu\nu}\cdot g\\
\Gamma^{\mu}=&\text{\parbox{25\unitlength}{
\includegraphics[scale=0.6]{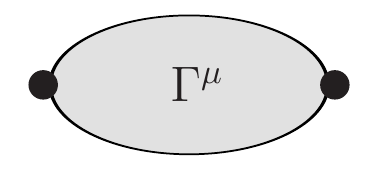}}}&
=w^{\mu}+g\cdot\Pi^{\mu\nu}\,w_\nu\,
,\nonumber
\label{correlation-structures}
\end{eqnarray}
which will provide standard random phase approximation RPA corrections
to the  selfenergies and vertices.  Here $w_\mu$  denotes the external
pion momentum.  The differences between these three expressions result
from the  outer most  vertices.  For $\Pi$  there are  two three-point
vertices at the outer  most positions, while $\Gamma^{\mu\nu}$ has two
four-point vertices  and $\Gamma^{\mu}$  one three and  one four-point
vertex.

For the  resulting selfenergies we  will omit contributions  which are
suppressed  by phase-space  constraints whenever  two ``simultaneous''
$\rho$-meson   lines  implicitly   occur  in   a  diagram.   The  pion
polarization function then becomes
\begin{eqnarray}
&&\Pi_{\pi}=\text{\parbox{44\unitlength}{\includegraphics[scale=0.6]{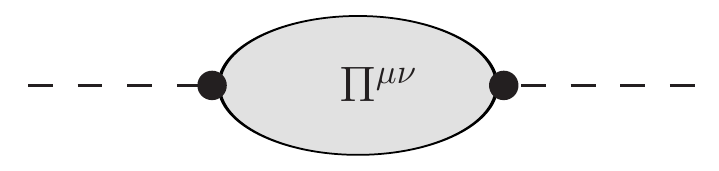}}}\,\nonumber\\
&&\qquad \left[+\,
\text{\parbox{45\unitlength}{\vspace{-0.2cm}\includegraphics[scale=0.5]{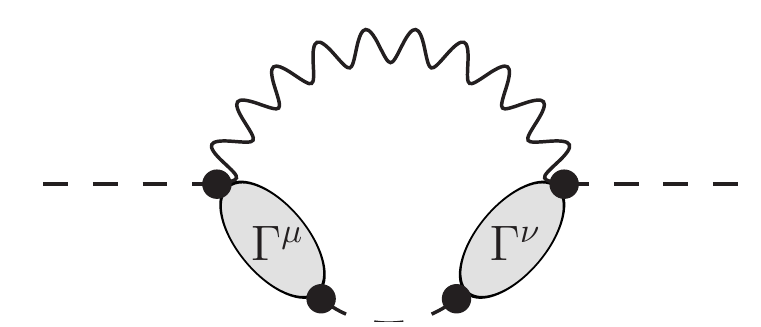}}}\,\right.\nonumber\\[6ex]
&&\qquad\left.+\,\text{\parbox{43\unitlength}{\vspace{-0.7cm}\includegraphics[scale=0.6]{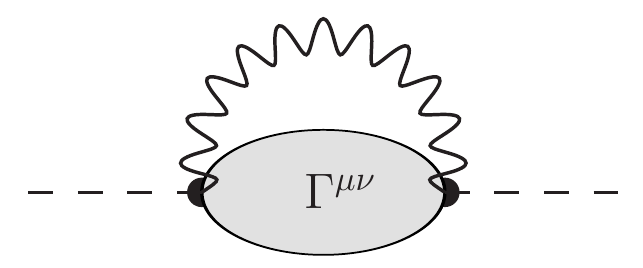}}}\right].\,\nonumber\\
&&=-4\,w^{\mu}\,\Pi_{\mu\nu}\,w^{\nu}+\delta m_{\pi}+w^2\,\delta,  
\label{Phi-S-Pi}
\end{eqnarray}
The  first  diagram gives  the  main  contribution.   It corrects  the
$\pi\rho$-loop in the pion  selfenergy by short-range correlations. The
two diagrams in brackets will  be omitted since they are suppressed by
phase-space constraints.   The renormalization terms  $\delta m_{\pi}$
and  $\delta$  in  (\ref{Phi-S-Pi})  will  be adjusted  in  vacuum  to
guarantee that the pion has its pole at $m^2=(139 \text{ MeV})^2$ with
residuum 1.  The polarization tensor of the $\rho$-meson is then given
by
\begin{eqnarray}
\Pi_{\rho}&=&\Pi_{(\rho,1)}+\Pi_{(\rho,2)}\qquad\text{with}\nonumber\\
-i\Pi_{(\rho,1)}&=&\text{\parbox{27\unitlength}{\vspace{-0.1cm}%
\includegraphics[scale=0.36]{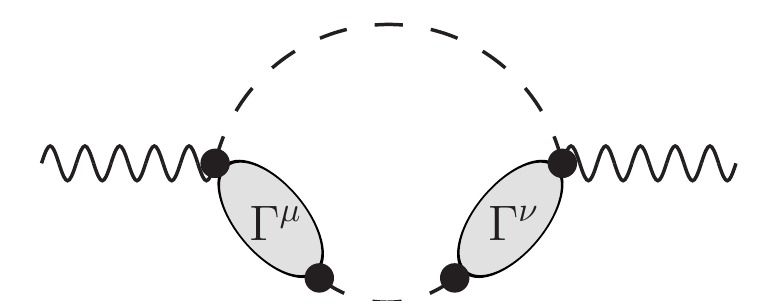}}}\, +
\text{\parbox{27\unitlength}{%
\includegraphics[scale=0.36]{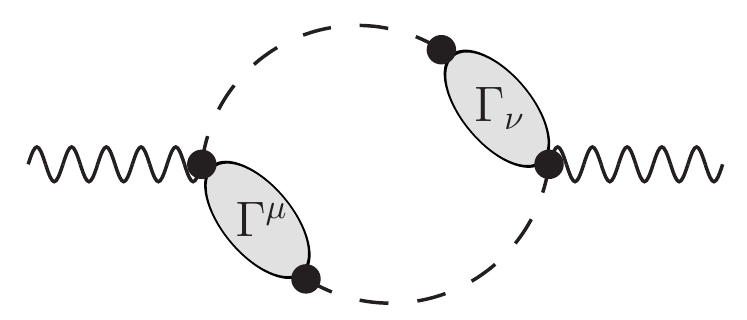}}}+\dots\nonumber\\
-i\Pi_{(\rho,2)}&=&\text{\parbox{30\unitlength}{{\vspace{+0.5cm}
\includegraphics[scale=0.4]{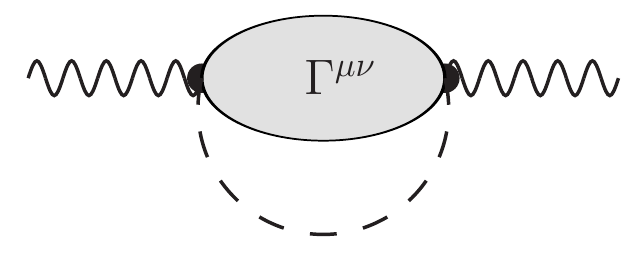}}}}\, ,
\label{rho-self}
\end{eqnarray}
where zero  or at  most one correlation  bubble $\Gamma^{\mu}$  can be
attached to the external  vertices of $\Pi_{(\rho,1)}$, the latter due
to the  $\pi\pi\rho\rho$ coupling of  the Lagrangian (\ref{L-vector}).
Again  phase-space suppressed  terms can  be dropped.   In the  end we
arrive at  a set of coupled  Dyson equations for  the determination of
the full retarded propagators in terms of the retarded selfenergies or
polarization  tensors and  the  free propagators.   Details about  the
calculation will be given in the next section. Readers only interested
in the  results could skip the  next section and directly  jump to the
result section.

%
\section{Details of the calculation}\label{SecDetailVec}
%

\subsection{Pion selfenergy and polarization loops\label{pion-self-energy-pi-rho}}

It   is  advantageous  to   decompose  the   central  correlation   loop
(\ref{correlation-loop}) into its Lorentz tensor components (see e.g.
\cite{Lutz:2002qy})
\begin{eqnarray}
& \chi_{\mu \nu}(w,u) =& \sum_{i,j=1}^2\,
\chi_{ij}(w,u)\,L_{\mu \nu}^{(ij)}(w,u)\nonumber\\
&&+ \chi_{\,T}(w,u)\,T_{\mu \nu}(w,u)\,.
\label{def-decom-pi-pi-rho}
\end{eqnarray}
Here $T_{\mu \nu}$ and $L_{\mu \nu}^{(22)}$ are the special projectors
on the two spatially  transverse and the spatially longitudinal modes.
The  other three  $L_{\mu  \nu}^{(ij)}$ complete  the tensor  algebra.
Furthermore $w$ and $u$ denote the external four momentum and the four
velocity of the equilibrated  matter, respectively (in the c.m.  frame
of the  matter $u=(1,\vec0\,)$). This decomposition  will simplify the
solution of  the Dyson equation (\ref{Dyson-pi-rho}) as  it provides a
decoupling   between   the   longitudinal  and   transversal   sectors
\cite{Lutz:2002qy}.   The derivation of  the explicit  expressions for
the components  $\chi_{ij}$ and $\chi_{\,T}$ is  relegated to Appendix
\ref{AppendixA}.  They  simply follow from contractions  of the tensor
$\chi_{\mu  \nu}(w,u)$   with  the  projectors.    The  decompositions
(\ref{def-decom-pi-pi-rho}) also  easily allow to  include the vertex
corrections.   We  first define  the  loop  matrices $\chi^{(L)}$  and
$\chi^{(T)}$
\begin{eqnarray}
&&\chi^{(L)}= \left(
\begin{array}{ll}
\chi_{11} & \chi_{12} \\
\chi_{21} & \chi_{22}
\end{array}
\right)\,
  \label{def-matrix-pi-rho} \,, \quad \!\chi^{(T)}= \left(
\begin{array}{l}
\chi_{T}
\end{array}
\right).
\label{chi-g-def-pi-rho}
\end{eqnarray}
The quantity $\Pi_{\mu\nu}(w,u)$, which sums up all correlations, then
results to
\begin{eqnarray}
&\Pi_{\mu\nu}(w,u)=&\sum_{i,j=1}^{2}\,\Pi_{(ij)}(w,u)\,
L_{\mu\nu}^{(ij)}(w,u)\nonumber\\  
&&+\Pi_{(T)}(w,u)T_{\mu\nu}(w,u)
\label{srt-pi-rho}
\end{eqnarray} 
with coefficient functions $\Pi_{ij}$ and $\Pi_{T}$ defined as
\begin{eqnarray}
&&\Pi_{(ij)}=\Big[\Big( \unitmatrix-\chi^{(L)}\,\Big)^{-1}
\chi^{(L)} \Big]_{ij}\nonumber\\
&&\Pi_{(T)}=\Big[\Big( \unitmatrix-\chi^{(T)}\,\Big)^{-1}
\chi^{(T)} \Big]\,.
\label{Def-R-pi-rho}
\end{eqnarray} 
Due to  the derivative structure of  the interactions (\ref{L-vector})
and   the   structure   of   the  four   particle   interactions   the
$\Gamma$-bubble    insertions    (\ref{correlation-structures})    and
(\ref{rho-self})  simply  lead  to  a  replacement of  the  bare  pion
momentum $w^{\mu}$ at the vertex by a dressed one
\begin{eqnarray}
&w_{\nu}\;\rightarrow\;\Gamma_\nu(w, u)&= 
w^\nu+ w^\mu\,\Gamma_{\mu\nu}(w,u)\\
&&=  w_\nu\,\Gamma_1(w,u) + u_\nu\,\Gamma_2(w,u)\,,\nonumber
\label{vertex-pi-rho}
\end{eqnarray} 
with contributions  proportional $w^{\mu}$ and $u^{\mu}$  given by the
vertex functions  $\Gamma_i$.  These vertex functions  are obtained by
contracting   the  full   correlation  sum   $\Pi_{\mu\nu}(w,u)$  over
$w^{\mu}$ because  one vertex directly  couples to the pion  while the
other one stems from the four point coupling. The two vertex functions
$\Gamma_1(q,u)$ and $\Gamma_2(q,u)$ are explicitly given by
\begin{eqnarray}
&\Gamma_{1}&=1+2\,\Big[\Big( \unitmatrix-g\,\chi^{(L)}\Big)^{-1}
g\,\chi^{(L)} \Big]_{11}\nonumber\\
&&+\frac{2\,(u\cdot w)}{\sqrt{w^2-(u\cdot w)^2}}
\Big[\Big( \unitmatrix-g\,\chi^{(L)}\Big)^{-1}
g\,\chi^{(L)} \Big]_{12}\nonumber\\
&&+\delta\Gamma\nonumber\\
&\Gamma_{2}&=\frac{-2\,w^2}{\sqrt{w^2-(u\cdot w)^2}}\,
\Big[\Big(\unitmatrix-g\,\chi^{(L)}\Big)^{-1}
g\,\chi^{(L)} \Big]_{12}\nonumber\\
\label{def-gamma-pi-rho}
\end{eqnarray} 
in  terms  of  the  loop  functions  (\ref{def-decom-pi-pi-rho}).   We
further introduced a finite renormalization $\delta\Gamma$ in order to
impose the  condition $\Gamma_{1}(w^2=m_{\pi}^2)=1$ in  vacuum.  There
are two important  technical issues to be emphasized  here. First, the
application   of  the  longitudinal   and  transverse   projectors  in
(\ref{def-decom-pi-pi-rho})  implies that the  loop functions  have to
satisfy specific  constraints. They  follow from the  observation that
the  polarization   tensor  $\chi_{\mu  \nu}(w,u)$   is  regular.   In
particular at $w^2 =0$ and at $w^2= (w \cdot u)^2$ it must hold that
\begin{eqnarray}
&&\chi_{22}(w,u)= \chi_{11}(w,u)-i\,\chi_{12}(w,u)-i\,
\chi_{21}(w,u)\nonumber\\
&&\qquad\qquad\qquad+ {\mathcal O} \left( w^2\right) 
\,,\nonumber\\
&& \chi_{22}(w,u)= 
\chi_T(w,u)+ {\mathcal O} \left( (w \cdot u)^2-w^2\right) \,.
\nonumber\\
\label{constraint-polarization}
\end{eqnarray}
These conditions  turn out  to be important  when specifying  the real
parts   of  the   loop  functions   (see   Appendix  \ref{AppendixA}).
Furthermore a finite renormalization  should be performed such that it
suppresses   the  formation   of   ghosts  in   the  pion   selfenergy
\cite{Korpa:2008ut}. The construction of the latter has also to comply
with the constraints (\ref{constraint-polarization}).

\subsection{Vector meson selfenergies}

Concerning the vector-meson polarization tensor special care has to be taken about two issues: Its  four transversality  and the  determination of its
regularized real part.   Let us start with the  four transversality. A
simple analysis  of the  Lorentz tensor decomposition  of $\Pi_{\rho}$
(\ref{rho-self})   into   the   projector   basis,  similar   as   for
(\ref{def-decom-pi-pi-rho}) or  (\ref{srt-pi-rho}), will directly show
that only in  the perturbative case no four  longitudinal modes arise.
The Dyson resummation (\ref{Dyson-pi-rho})  will lead to non vanishing
four-longitudinal  components, i.e.   $\Pi^{(11)}_{(\rho)}\neq  0$ and
$\Pi^{(12)}_{(\rho)}=\Pi^{(12)}_{(\rho)}\neq    0$.     This   problem
originates from the violation of Ward-identities in the selfconsistent
treatment.  Several  schemes were proposed in literature  to cure this
problem     \cite{vanHees:2000bp,     Riek:2004kx,     Ruppert:2004yg,
  Riek:2006vq, RR-Eratum}\footnote{A further possibility to circumvent
  this  problem  is  given  by  the tensor
  representation  of vector  mesons  \cite{Leupold:2006bp}.  Then  the
  propagation  of  four-longitudinal modes  is  not  supported by  the
  structure  of the  vertices.  Here,  however, we  stick to  the more
  common vector representation.}  All  schemes rely on some projection
procedure
\begin{eqnarray}
\Pi^{(ij)}_{(\rho)},\,\Pi^{(T)}_{(\rho)}\,\longrightarrow\,
\Pi^{(22)}_{(\rho,P)},\,\Pi^{(T)}_{(\rho,P)} 
\end{eqnarray}
where  from the  selfconsistently calculated  $\Pi_{(\rho)}$  with its
coefficients  on  the  left  a  fully  four-transversal  structure  of
$\Pi_{(\rho,P)}$   is  determined.    However,  as   pointed   out  in
\cite{Riek:2006vq}  all  schemes  used  so  far violate  some  of  the
constraints (\ref{constraint-polarization})  and therefore suffer from
the   occurrence   of   kinematical   singularities.   As   shown   in
\cite{Riek:2006vq} such singularities  have a substantial influence on
the calculation.   Here we  will follow the  scheme introduced  by van
Hees and Knoll \cite{vanHees:2000bp} and show how it could be modified
to  avoid this  problem.   This scheme  respects particular  dynamical
properties of the polarization tensor.   It exploits the fact that the
spatial components of the polarization tensor $\Pi^{ik}_{(\rho)}$ have
a finite relaxation time and are of no particular harm.  Thus they can
be kept $\Pi_{(\rho,P)}^{ik}=\Pi_{(\rho)}^{ik}$.  The time-components,
however,  involve an infinite  relaxation time,  since they  carry the
information about  the conservation  laws.  Such components  can never
reliably be  calculated at finite  loop order.  These  time components
can  however be constructed  solely from  the spatial  components such
that  the  full tensor  becomes  four-transversal.   Thus, the  scalar
functions  $\Pi^{(22)}_{(\rho,P)}$ and  $\Pi^{(T)}_{(\rho,P)}$  of the
three physical modes, the  spatially longitudinal and transverse ones,
are calculated solely from  the spatial parts of the polarization
tensors using the following spatial traces
\begin{eqnarray}
   \Pi_1=\!&\frac{w_iw_k}{\vec{w}^{\,2}}\Pi^{ik}_{(\rho)};\qquad\quad
   3\Pi_3&\!=-g_{ik}\Pi^{ik}_{(\rho)}\\ 
   \Pi_{(\rho,\,P)}^{(22)}=\!&\frac{w^2}{(u\cdot w)^2}\cdot\Pi_1;\quad\;
   \Pi_{(\rho,\,P)}^{(T)}&\!=\frac{1}{2}\left(3\Pi_3-\Pi_1\right).\quad
   \label{Spur}
\end{eqnarray}
Therefore  this scheme  has  a physically  sound background.   However
unless $\Pi_1$ vanishes quadratically towards zero energy $w^0=(u\cdot
w)$,  which generally  will  not  be the  case,  a singularity  occurs
\cite{Riek:2006vq}. Placed in  the space-like region the corresponding
spurious  zero   energy  mode   does  not  directly   affect  physical
observables such  as dilepton spectra.  It will  however influence the
selfconsistent dynamics,  if the  coupling of vector-mesons  back onto
other  particles in  the  system is  considered\footnote{Note that  in
  \cite{Riek:2004kx}  where we  used  this scheme  the propagation  of
  spurious   modes  was   blocked  due   to  the   structure   of  the
  $\pi\omega\rho$-vertex.}.

The advantage  of this scheme is  that it is free  of singularities in
the entire  time-like region.  It  therefore opens the  perspective to
construct a singularity free tensor by some infrared cut-off procedure
solely    applied    to    the    spatial    longitudinal    component
$\Pi_{(\rho,\,P)}^{(22)}$ in the  space-like region close to vanishing
energy.   To  do so  we  rewrite  the  relation for  the  longitudinal
projector (\ref{Spur}) as
\begin{eqnarray}
  &\Pi_{(\rho,\,P)}^{(22)}=&\Pi^{(22)}_{(\rho)}
  -\frac{(u\cdot w)^2-w^2}{(u\cdot w)^2}\,\Pi^{(11)}_{(\rho)}\nonumber\\
  &&+2i\,\frac{\sqrt{(u\cdot w)^2-w^2}}{(u\cdot w)}\,\Pi^{(12)}_{(\rho)},
\label{Spur2}
\end{eqnarray} 
where  we   used  $\Pi^{(21)}_{(\rho)}=\Pi^{(12)}_{(\rho)}$.  In  this
formulation  we directly  see that  (\ref{constraint-polarization}) is
perfectly reproduced  on the light cone  so the selfenergy  is free of
singularities there.  The same is true at  vanishing spatial momentum.
The   singularities    stem   from    the   factors   in    front   of
$\Pi^{(11)}_{(\rho)}$  and  $\Pi^{(12)}_{(\rho)}$  at $(u\cdot  w)=0$.
Thus one  can attempt  to construct the  $\Pi^{(22)}_{(\rho,\,P)}$ and
$\Pi^{(T)}_{(\rho,\,P)}$ coefficients as
\begin{eqnarray}
&&\Pi^{(T)}_{(\rho,\,P)}=\Pi_{(\rho)}^{(T)}\\
&&\Pi^{(22)}_{(\rho,\,P)}=\Pi^{(22)}_{(\rho)}
-\lambda(w,u)\,\Pi^{(11)}_{(\rho)}-2i\,
\sqrt{\lambda(w,u)}\,\Pi^{(12)}_{(\rho)}\nonumber
\label{P-coeff}
\end{eqnarray} 
with a coefficient function $\lambda(w,u)$, which has to fulfill
\begin{eqnarray}
&& \lambda(w^2=0,u)=1
\label{G-conditions}
\end{eqnarray} 
and should stay finite towards  $(u\cdot w)=0$. A possible choice that
provides a smooth transition to the form (\ref{Spur2}), which we would
like to keep due to its physical motivation is given by
\begin{eqnarray}
\hspace*{-0.4cm}\lambda(w,u)\!&=&\!\!\left\{\!\!
\begin{array}{ll}
\frac{(u\cdot w)^2-w^2+\Lambda^2}{2((u\cdot w)^2
+\Lambda^2)}+\frac{(u\cdot w)^2-w^2}{2(u\cdot w)^2 -w^2}&
  \mbox{for }w^2<0\\[2mm]
\frac{(u\cdot w)^2-w^2}{(u\cdot w)^2}&\mbox{for }w^2>0.
\end{array}
\right.
\label{interpolation-scheme}
\end{eqnarray} 
Here the parameter $\Lambda$  regularizes the infrared singularity and
controls the strength  in the far space like  region. Later variations
of $\Lambda$ can then be used to control the uncertainty introduced by
the cut-off.

We now  now turn to  the determination of  the real parts.   Since the
imaginary parts of the loops do  not drop to zero for large energies renormalization is required which we
introduce using subtracted dispersion relations.  Thereby one  has to keep
in mind that  along with the imaginary parts also  the real parts have
to  be  free  from   kinematical  singularities  and  thus have to  fulfil
(\ref{constraint-polarization}).

At the vacuum level it suffices to consider the following subtracted
dispersion relation 
\begin{eqnarray}
\Pi^{(22/T)}_{(\rho,\,P)}(w)=\frac{1}{\pi}\,
\int d\bar{w}_0\,\frac{w^4}{\bar{w}^4}
\frac{\Im\,\Pi_{(\rho,\,P)}^{(22/T)}(\bar{w})}{w_0-\bar{w}_0+i\epsilon}
\label{rho-realpart-vac}
\end{eqnarray} 
with  $w=(w_0,\vec{w}\,)$   and  $\bar{w}=(\bar{w}_0,\vec{w}\,)$.   It
automatically  guarantees   that  the  polarization   tensor  and  its
derivative  vanish   on  the  light  cone  so   that  the  kinematical
constraints  (\ref{constraint-polarization}) are  naturally fulfilled.
This  technically  preferred  renormalization  guarantees  a  massless
photon  with pole  residuum  1 within  the  vector dominance  picture.
However, once medium effects come  into play and all thresholds become
effectively removed,  the imaginary parts  do no longer vanish  on the
light  cone. One method to  extend the  prescription to  the in-medium
situation, is to first convert  the description  to a
singularity free  basis, then  perform the dispersion  integrals which
are  then free  of  any constraints  and  subsequently reconvert  back
\cite{Phd-Riek} to the tensor decomposition.  After combining all this
together with the vacuum prescription we obtain
\begin{eqnarray}
&&\Pi^{(T)}_{(\rho,\,P)}(w,u)=\Pi^{(T)}_{(\rho,vac)}(w,u)\nonumber\\
&&+\int\frac{(u\cdot w)^2}{( u\cdot\bar{w})^2}
\frac{\Im\,\Pi^{(T)}_{(\rho,\,P)}(\bar{w},u)
-\Im\,\Pi^{(T)}_{(\rho,vac)}(\bar{w},u)}
{\bar{w}_0-w_0+i\epsilon }\,d\bar{w}_0\nonumber\\[1.3 ex]
&&\Pi^{(22)}_{(\rho,\,P)}(w,u)=
\Pi^{(22)}_{(\rho,vac)}(w,u)\nonumber\\
&&+\int\frac{w^2}{{\bar w}^2}
\frac{\Im\,\Pi^{(22)}_{(\rho,\,P)}(\bar{w},u)
-\Im\,\Pi_{(\rho,vac)}^{(22)}(\bar{w},u)}
{\bar{w}_0-w_0+i\epsilon}\,d\bar{w}_0\,.\nonumber\\
\label{Realparts-Full}
\end{eqnarray} 
Here it  is understood that the  vacuum terms $\Pi^{(T)}_{(\rho,vac)}$
and  $\Pi^{(22)}_{(\rho,vac)}$  already   contain  the  projection  to
restore  four  transversality.   The  factor $w^2$  in  the  spatially
longitudinal  term  is essential  to  cancel  the $1/w^2$  singularity
arising  from  the  projector.   The prescription  then  automatically
guaranties  that the  three longitudinal  and three  transversal parts
become degenerate for zero spatial momentum.
\begin{figure}[b]
\includegraphics[scale=0.7]{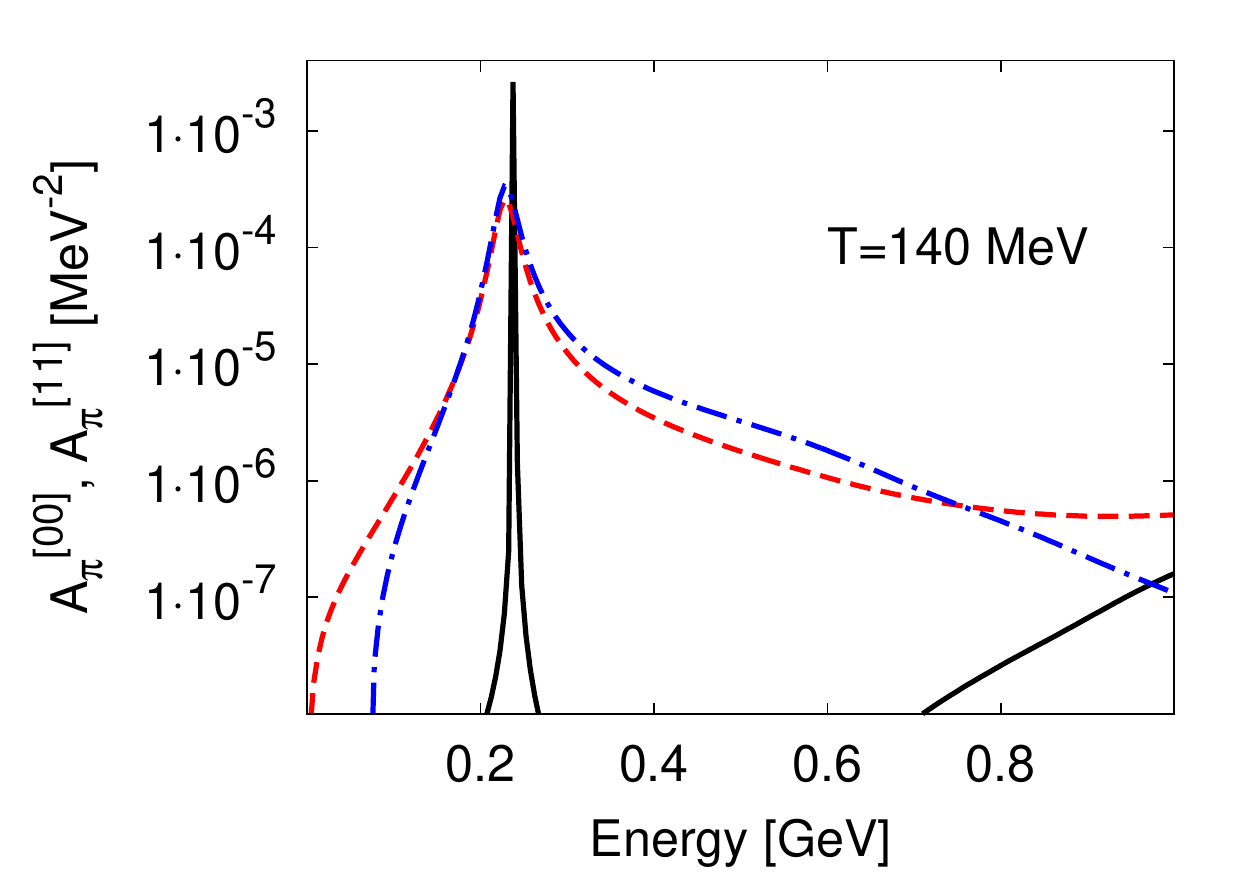}
\caption{Pion spectral function $A^{[00]}_{\pi}$ (dashed line) and the
  effective spectral function $A_{\pi}^{[11]}$ (dashed dotted line) at 
  $T=140$~MeV compared to the vacuum spectral function (full line) for
  a momentum of 200 MeV.\label{Api-plot}}
\end{figure}

Now we comment about the inclusion of vertex corrections. They can be
included along the lines of Refs. \cite{Korpa:2008ut,Phd-Riek} by
introducing effective spectral functions which include the vertex
structure (\ref{vertex-pi-rho})
\begin{eqnarray}
A_{\pi}^{[ij]}(q,u)=-2\,\Im\left[
\frac{\Gamma_{i}(q,u)\,\Gamma_{j}(q,u)}
{q^2-m_{\pi}^{2}-\Pi_{\pi}(q,u)}\right]\,. 
\label{eff-A}
\end{eqnarray} 
We  further define  $\Gamma_{0}(q)=1$  which allows  us  to write  the
normal  pion spectral  function as  $A_{\pi}^{[00]}(q,u)$.  Collecting
then the vertex tensors  $\Gamma_{i}$ into the pion spectral functions
as defined in (\ref{eff-A}) the expressions for the imaginary parts of
the   selfenergies  (\ref{rho-self})   can  straight   forwardly  be
evaluated.   The results  can  be found  in Appendix  \ref{AppendixB}.
From these  the complete polarization tensors can  be calculated using
(\ref{P-coeff}) and (\ref{Realparts-Full}).

\section{Results\label{results}}
%

As the  main focus of our  work is on the  conceptual developments, we
keep  this section rather  brief, concentrating  on the  most relevant
results only.  First we analyze the influence of the cut-off $\Lambda$
introduced through the projection scheme (\ref{interpolation-scheme}).
We found only  a small sensitivity on the  choice of the interpolation
and   therefore   use   a   value   of  $\Lambda=200$   MeV   in   the
following\footnote{However  any attempt to  use $\Lambda=0$  would, as
  expected, fail completely.}. This is good news because it shows that
as soon as these space-like modes are treated properly their influence
is rather small and the  treatment with some a priori unknown infrared
cut-off does not introduce a large uncertainty in the calculations.

From analytic estimates and earlier calculations \cite{Riek:2004kx} we
expect no dramatic changes of the spectral distribution of the
$\rho$-meson. The most interesting point will be what influence the
vertex corrections have on the result and to what extend the pion gets
modified through selfconsistency.

In  Fig.   \ref{Api-plot}  the  resulting pion  spectral  function  is
presented for zero and 140 MeV temperature.  As compared to the vacuum
case we observe  that in the medium the gap  between the on-shell pole
and the $\rho\pi$-continuum becomes  filled and that the on-shell peak
gets  broadened.   In addition  low  energy  components  arise due  to
scattering off thermal pions.  The effect of the vertex correction can
be  seen   when  comparing  $A^{[00]}_{\pi}$   with  $A_{\pi}^{[11]}$.
Since  $\Gamma_1$ is  complex,  also  the real  part  of the  pion
  propagator contributes to $A_{\pi}^{[11]}$.  Since far away from the
  pole  the real  part  is much  larger  than the  imaginary part  and
  changes  sign   at  the  pion   pole,  one  obtains   a  destructive
  interference  at low  energies and  some enhancement  in  the region
  between the  on-shell pole  and the continuum.   This shift  in pion
  strength leads to a  reduced broadening of the $\rho$-meson as
compared to the case without vertex correction because the phase space
for the decay becomes reduced.  However the influence of the vertex is
much   smaller   then   in    the   case   of   baryonic   excitations
\cite{Korpa:2008ut} as could already  be expected from the rather high
threshold of the $\rho\pi$-loop as  compared to the pion mass. Effects
arising from  the other components of the  effective spectral function
are zero in vacuum and stay negligible in the medium so that we do not
discuss them here.
\begin{figure}[t]
\includegraphics[scale=0.7]{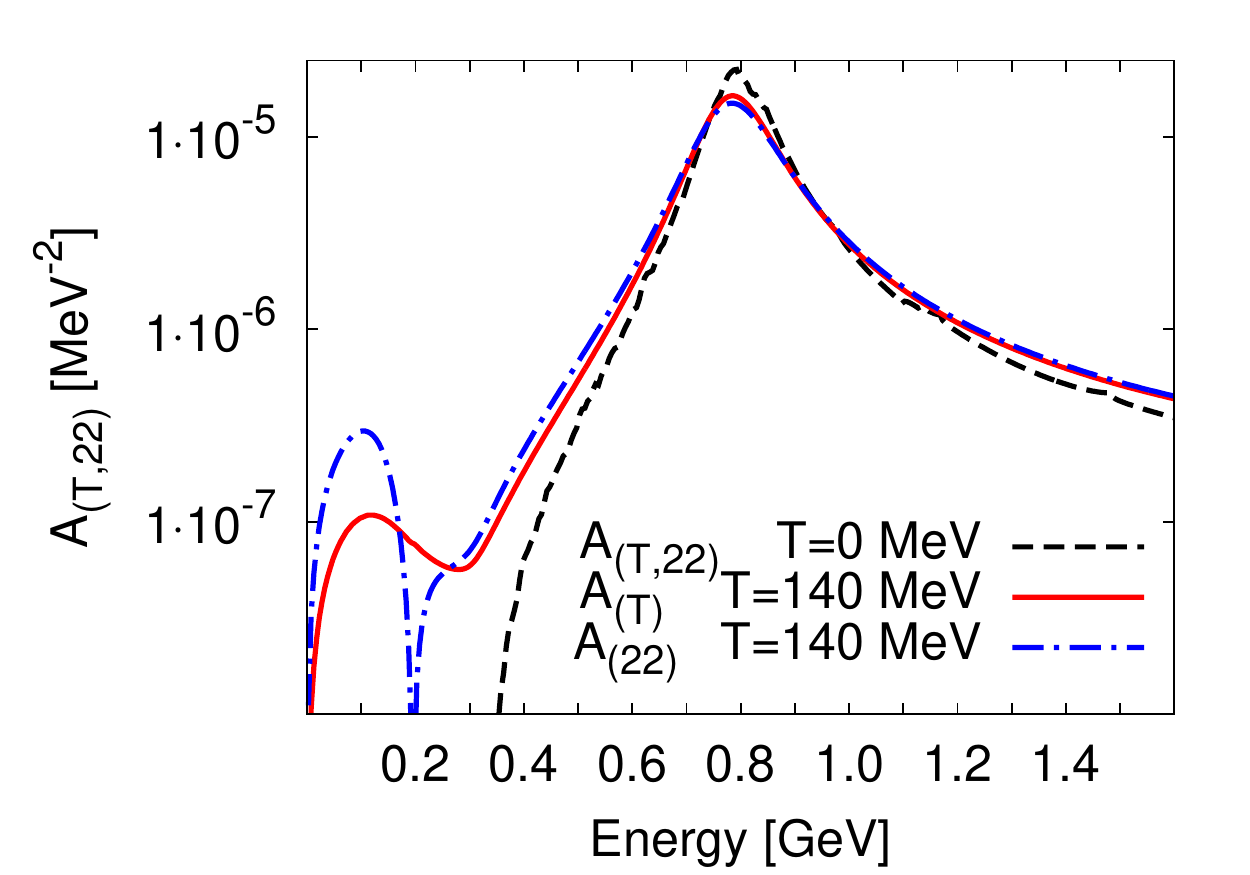}
\caption{Spectral  functions   of  the  $\rho$-meson   at  $T=140$~MeV
  compared to the vacuum spectral  function.  The momentum is 200 MeV.
  For  the  spatially longitudinal  spectral  function $A_{(22)}$  the
  values   are   negative   below   the  light   cone,   i.e.    below
  200~MeV.\label{Arho-plot}}
\end{figure}

\begin{figure}
\includegraphics[scale=0.54]{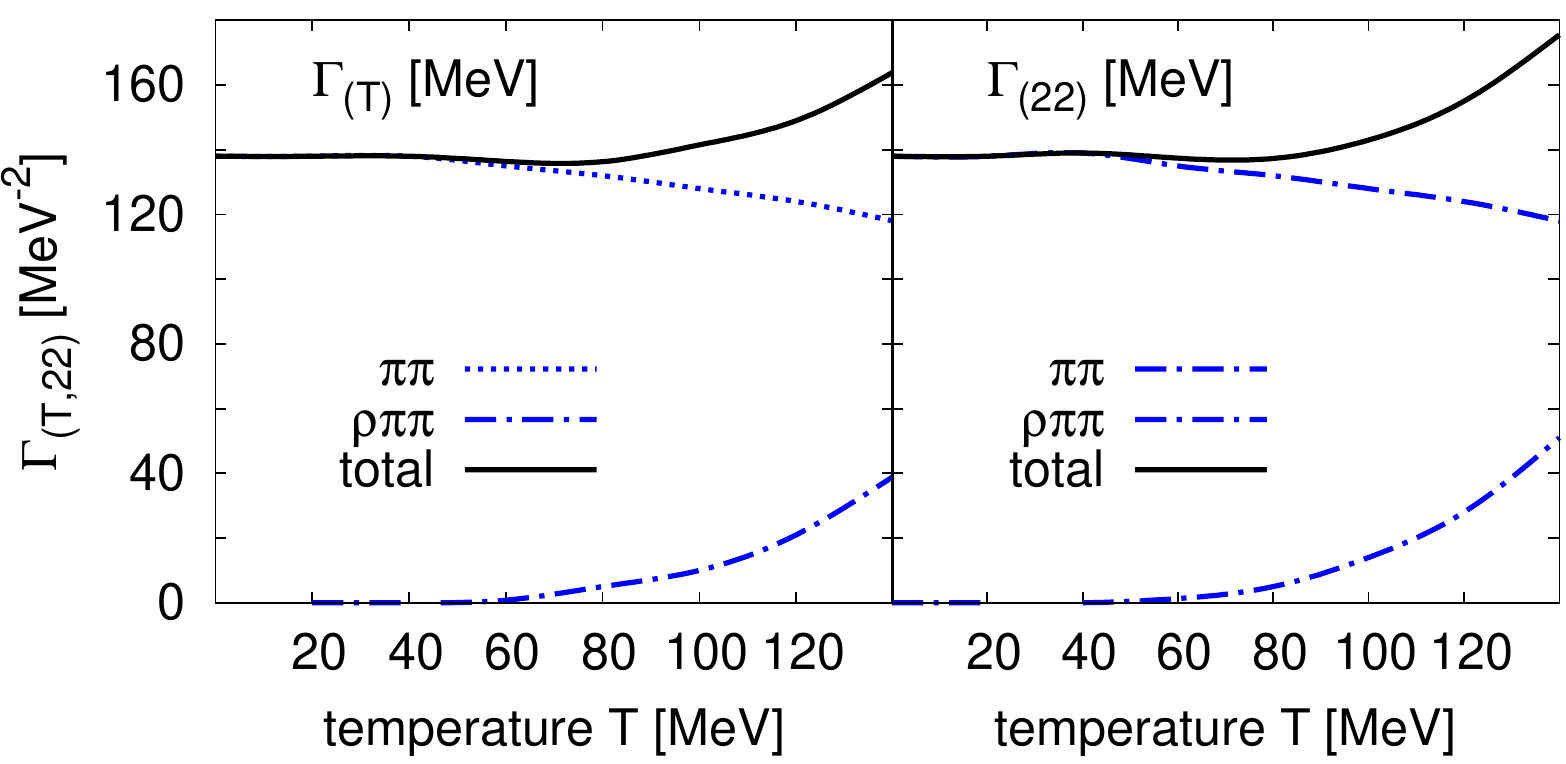}
\caption{Spatially transverse  and longitudinal damping  widths of the
  $\rho$-meson  versus  temperature  divided  into  the  two  dominant
  partial channels.}\label{rho-meson-width-plot}
\end{figure}

The   $\rho$-meson  spectral  function   shows  only   minor  changes.
Noteworthy is that  the partial width resulting from  the decay into a
pion   pair  becomes   reduced  at   higher  temperatures   (see  Fig.
\ref{Arho-plot} and \ref{rho-meson-width-plot}). This is caused by the
asymmetry of  the pion  spectral function around  the pole  mass which
receives a larger strength on the  high mass side due to the $\rho\pi$
cut   and  therefore   kinematically  disfavors   the  decay   of  the
$\rho$-meson. The resulting net  effect between this reduction and the
thermal enhancement turns  out to be quite small  such that the actual
enhancement of the $\rho$-meson width  mainly stems from the new decay
mode    into   $\rho\pi\pi$    channel    (cf.    $\Pi_{\rho,2}$    in
(\ref{rho-self})) which has a lower threshold in dense matter once
all particles attain broad  spectral distributions.  However even with
the   additional   scattering  and   decay   possibilities  into   the
$\rho\pi\pi$ channel the width is  only marginally increased by 35 MeV
at 140 MeV temperature (see Fig.  \ref{rho-meson-width-plot}).  In all
cases the momentum dependence proved to be small.

%
\section{Conclusions}
%

In this  work we developed  some technical concepts for  the in-medium
treatment of  vector mesons in a  selfconsistent framework.  Therefore
we studied the influence of the presence of hot matter on the spectral
properties of  the $\rho$  meson. The aim  was to  include short-range
correlations of  the Migdal type in  order to consistently  sum up all
soft  modes of the  system. These  short-range correlations  which are
normally used  to describe the  interactions of the pion  with nucleon
and $\Delta$-isobar  were now also  applied to the vector-mesons  in a
selfconsistent   framework.    Special  emphasis   was   put  on   the
determination  of the real-parts  of all  selfenergies and  the proper
avoiding  of kinematical singularities  in the  selfconsistent scheme.
The treatment of vector-mesons within the current model setup requires
great care  due to the fact  that the polarization tensors  have to be
kept four transversal in order  to avoid the propagation of unphysical
degrees of freedom.

In the  purely mesonic  system containing pions  and $\rho$  mesons no
large medium  effects were obtained, neither at  high temperatures nor
due to  the here considered correlations and  vertex corrections. This
complies   with   earlier   studies   \cite{Rapp:1999ej,Peters:1997va,
  Post:2000qi,Herrmann:1993za,Friman:1997tc}, where  it was found that
the  dominating in-medium  effects on  the light  vector-mesons result
from the direct  interaction with baryons. We only  observe a moderate
broadening  of about 30  MeV for  both vector-mesons  even at  140 MeV
temperature.   However,  compared to  the  perturbative treatment  the
selfconsistent  scheme suggest  a  further source  of the  broadening,
namely the $\rho\pi\pi$ decay. Of  course this picture will be greatly
influenced by the presence  of low energy particle-hole excitations in
the pion channel  possibly leading to different conclusions  in a more
complete  model.  The influence  of the  vertex corrections  and short
range  correlations turned  out  to  be quite  small  compared to  the
over-all size of  the selfenergies. On the other  hand interesting new
effects  have  been  found  which  are  necessary  to  understand  the
microscopic interactions in more detail.  The influence of both vertex
corrections and selfconsistency is  expected to become important, once
the scheme is extended to include the direct coupling to baryons. Then
new   low   energy   resonance-hole   excitations   come   into   play
\cite{Korpa:2008ut}. Using the formalism developed in this work and in
\cite{Korpa:2008ut} such extensions can be addressed in the future.

\vspace{0.5cm}

\textbf{Acknowledgments}\\

The authors acknowledge fruitful discussions with M.F.M. Lutz, B.
Friman, R. Rapp, S. Leupold, D. Rischke, J. Ruppert and D.
Voskresensky at various stages of this work. FR was supported by 
U.S. NSF grants PHY-0449489 (CAREER) and PHY-0969394.

\appendix

\section{$\pi\rho$ loop tensor coefficients\label{AppendixA}}

The correlation loop (\ref{correlation-loop}) is defined as:
\begin{eqnarray}
&\mathbf{\chi}^{\mu\nu}&=
8g^2\int\frac{d^4l}{2(2\pi)^4}G_{\mu\nu}(l,u)\,D(l-w,u)\,.
\end{eqnarray}
The $\pi\rho$ loop contains an isospin factor of two due to isospin
symmetry.  In contrast to the dispersion relation strategy employed
for the vector-meson, we here use a formfactor
\begin{eqnarray}
F(q^2)=\left(\exp\left(\frac{w^2-\lambda^2}{\lambda^2}\right)\right)^2
\theta(w^2-\lambda^2)+\theta(\lambda^2-w^2)\nonumber\\
\end{eqnarray} 
with $\lambda=1250$ MeV, since some hard scale could be chosen without
problems.   The  imaginary  parts  of  the  $\rho\pi$  loop  functions
$\chi_{ij}$ of Eq. (\ref{def-decom-pi-pi-rho}) are then given by
\begin{widetext}
\begin{eqnarray}
&&\Im\chi_{ij}(w,u)=2g\int\frac{d^4l}{(2\pi)^4}\,(H^{[ij,22]}\,
A^{\rho}_{22}(l,u)+H^{[ij,T]}\,A^{\rho}_{T}(l,u))\, 
A^{[00]}_{\pi}(l-w,u)\,(n_{B}(l\cdot u)-n_{B}((l-w)\cdot
u))\,F(w^2)\,,\nonumber\\
&&\Im\chi_{T}(w,u)=2g\int\frac{d^4l}{(2\pi)^4}\,(H^{[T,22]}\,
A^{\rho}_{22}(l,u)+H^{[T,T]}\,A^{\rho}_{T}(l,u))\, 
A^{[00]}_{\pi}(l-w,u)\,(n_{B}(l\cdot u)-n_{B}((l-w)\cdot
u))\,F(w^2)\,,\nonumber
\end{eqnarray}
in terms  of coefficients $H^{[11,22]}$ and  $H^{[11,T]}$ specified at
the end of this  section (\ref{H-coeff}) and the $\rho$-meson spectral
function $A^{\rho}$ which has also been decomposed using the projector
algebra.   In addition  we have  to  take care  about the  kinematical
constraints.   This  can be  realized  along  the  lines presented  in
\cite{Korpa:2008ut}  for  baryonic   loops  by  choosing  a  different
representation
\begin{eqnarray}
&&\chi_{11} (\omega, \vec q\,) = \frac{1}{q^2}\,\chi_1 (\omega, \vec
q\,)\,,\qquad 
\chi_{12} (\omega, \vec q\,) =
\frac{1}{\sqrt{q^2-(q\cdot u)^2}}\Bigg(\frac{q \cdot u}{q^2}\,
\chi_1 (\omega, \vec q\,)- \chi_2 (\omega, \vec q\,) \Bigg)\,,
\\
&& \chi_{22} (\omega, \vec q\,) =\frac{q \cdot u}{q^2-(q \cdot u)^2}\,\Bigg(
 \frac{q^2}{q\cdot u}\,\chi_3 (\omega, \vec q\,)-2\,\chi_{2} (\omega,
 \vec q\,) \Bigg)\,,\quad 
 \chi_{T} (\omega, \vec q\,) = \frac{1}{2}\,\left(
\chi_4 (\omega, \vec q\,)-\chi_{11} (\omega, \vec q\,)-\chi_{22}
(\omega, \vec q\,)\right)\,.\nonumber 
\label{rewrite-pirho}
\end{eqnarray}
The new functions $\chi_{i}$ can now be obtained, using the kernels
defined in (\ref{H-coeff}),
\begin{eqnarray}
&& \chi_{i} (\omega , \vec q\,)= \Bigg[\delta_{i4}\,\chi_3(0, \vec q\,)-2g
\int\frac{d^4l}{(2\pi)^4} \int_{-\infty }^{+\infty } \frac{d \bar
  \omega }{\pi}\,\left( \frac{q^2 }{\bar q^2}\right)^{n_i}\, 
\frac{(H^{[i,22]}\,A^{\rho}_{22}(l,u)+H^{[i,T]}\,A^{\rho}_{T}(l,u))}
{\bar \omega - \omega- i\,\epsilon}
\nonumber\\
&&\qquad \qquad \qquad \times A^{[00]}_{\pi}(l-w,u)\,(n_{B}(l\cdot
u)-n_{B}((l-w)\cdot u))\,F(w^2) \Bigg]\quad  \;+ (q_\mu \to -q_\mu
)\,,\nonumber 
\end{eqnarray}
for $i=1,3,4$ with $n_{1,4}=2$, $n_{2}=1$ and
$q^2=\omega^2-\vec{q}^{\,2}$,
$\bar{q}^2=\bar{\omega}^2-\vec{q}^{\,2}$. While for $n=2$ we have
\begin{eqnarray}
&& \chi_{i} (\omega , \vec q\,)= \Bigg[\delta_{i4}\,\chi_3(0, \vec q\,)-2g
\int\frac{d^4l}{(2\pi)^4} \int_{-\infty }^{+\infty } \frac{d \bar
  \omega }{\pi}\,\left( \frac{\omega}{\bar \omega}\right)\, 
\frac{(H^{[i,22]}\,A^{\rho}_{22}(l,u)+H^{[i,T]}\,A^{\rho}_{T}(l,u))}
{\bar \omega - \omega- i\,\epsilon}
 \nonumber\\
&& \qquad \qquad \qquad \times A^{[00]}_{\pi}(l-w,u)\,(n_{B}(l\cdot
u)-n_{B}((l-w)\cdot u))\,F(w^2) \Bigg]\;-(q_\mu \to -q_\mu
)\,,\nonumber 
\end{eqnarray}
\end{widetext}
It remains to specify the coefficients $H^{[11,22]}$ and $H^{[11,T]}$
(we use $y\in\{ij,T\}$): 
\begin{eqnarray}
&&
H^{[T,lm]}=\frac{1}{2}g^{\nu\alpha}g^{\mu\beta}\,T_{\mu\nu}(w,u)\,
L^{(ij)}_{\alpha\beta}(l,u)\nonumber\\  
&& H^{[ij,T]}=g^{\nu\alpha}g^{\mu\beta}\,L^{(ij)}_{\mu\nu}(w,u)\,
T_{\alpha\beta}(l,u) 
\label{H-coeff}
\end{eqnarray} 
\begin{eqnarray}
&& H^{[ij,lm]}=g^{\nu\alpha}g^{\mu\beta}\,L^{(ij)}_{\mu\nu}(w,u)\,
L^{(ij)}_{\alpha\beta}(l,u)\nonumber\\
&& H^{[1,11]}=\frac{\ql^2}{l^2}\qquad H^{[2,11]}=
\frac{\ul \, \ql}{l^2}\nonumber\\
&& H^{[3,11]}=\frac{\ul^2}{l^2}\qquad H^{[4,11]}=1\nonumber
\end{eqnarray} 
\begin{eqnarray}
&& H^{[11,y]}=\frac{1}{w^2}\,H^{[1,y]},\nonumber\\ 
&& H^{[12,y]}=\frac{1}{\sqrt{w^2-\uq^2}}\left[\frac{\uq}{w^2}\,
H^{[1,y]}-H^{[2,y]}\right]\nonumber\\
&& H^{[22,y]}=\frac{\uq}{w^2-\uq^2}\left[\frac{\uq}{w^2}\,
H^{[1,y]}\right.\nonumber\\
&& \qquad\qquad\qquad\left.-2\,H^{[2,y]}+\frac{w^2}{\uq}\,
H^{[3,y]}\right],\nonumber\\ 
&& H^{[T,y]}=\frac{1}{2}\left[H^{[4,y]}-H^{[11,y]}-H^{[22,y]}\right]\nonumber
\end{eqnarray} 
\begin{eqnarray}
&& H^{[1,12]}=\frac{\ql\,(\ul\,\ql-l^2\,\uq)}{l^2\sqrt{l^2-\ul^2}}\nonumber\\
&& H^{[2,12]}=-\frac{\ql\,\sqrt{l^2-\ul^2}}{l^2}\nonumber\\
&& H^{[3,12]}=\frac{1}{l^2\,\sqrt{l^2-\ul^2}}\left[w^2\,\ul^3\right.\nonumber\\
&& \qquad\qquad\left.-l^2\,(\uq\ql+\ul\,(w^2-\uq^2))\right]\nonumber\\
&& H^{[4,12]}=0\nonumber
\end{eqnarray} 
\begin{eqnarray}
&& H^{[1,22]}=\frac{(\ul\,\ql-l^2\,\uq)^2}{l^2\,(l^2-\ul^2)}\nonumber\\
&& H^{[2,22]}=\uq-\frac{\ql\,\ul}{l^2}\nonumber\\
&& H^{[3,22]}=1-\frac{\ul^2}{l^2}\qquad H^{[4,22]}=1\nonumber
\end{eqnarray} 
\begin{eqnarray}
&& H^{[1,T]}=\frac{-1}{2\,(l^2-\ul^2)}\left[\ql^2+\ul^2\,w^2\right.\nonumber\\
&& \qquad\qquad\left.-2\,\uq\,\ul\,\ql+l^2\,(\uq^2-w^2)\right]\nonumber\\
&& H^{[2,T]}=0\qquad H^{[3,T]}=0\qquad H^{[4,T]}=1\nonumber
\end{eqnarray} 
\begin{eqnarray}
& H^{[T,21]}=H^{[T,12]}\qquad &H^{[22,21]}=H^{[22,12]}\nonumber\\
& H^{[11,21]}=H^{[11,12]}\qquad &H^{[12,21]}=H^{[21,12]}\nonumber\\
& H^{[21,21]}=H^{[12,12]}\qquad &H^{[21,22]}=H^{[12,22]}\nonumber\\
& H^{[21,T]}=H^{[12,T]}\qquad &H^{[21,11]}=H^{[12,11]}\nonumber
\end{eqnarray}

\subsection{Coefficients of the vector-meson selfenergies\label{AppendixB}}

We  calculate  the  expressions  for  the  vector-meson  selfenergies.
According to (\ref{rho-self}) the polarization tensors are given by
\begin{eqnarray}
&&\Pi^{\mu\nu}_{(\rho,1)}(w,u) = 
g^{2}\int \frac{d^4l}{2(2\pi)^4}\,\Big[(\Gamma^{\mu}(l,u)
+\Gamma^{\mu}(l-w,u))\nonumber\\
&&\,(\Gamma^{\nu}(l,u)
+\Gamma^{\nu}(l-w,u))\,G_{\pi}(l,u)\,G_{\pi}(l-w,u)\Big]\nonumber\\
&&\Pi^{\mu\nu}_{(\rho,2)}(w,u)=
g^2\int \frac{d^4l}{2(2\pi)^4}\Pi^{\mu\nu}(l,u)\,G_{\pi}(l+w,u)\,.
\nonumber\\
\label{S-rho-diagram} 
\end{eqnarray}
These  expressions have to be  decomposed  into the  coefficient
functions  $\Pi^{(ij)}_{(\rho,i)}$ and  $\Pi^{(T)}_{(\rho,i)}$.  Using
the   functions  $B$   and  $H$   specified  in   (\ref{B-coeff})  and
(\ref{H-coeff})  and the  functions  $\Pi_{ij}$ which  are defined  in
(\ref{Def-R-pi-rho})   and   taking   the  pion   spectral   functions
$A^{[ij]}_{\pi}$ from (\ref{eff-A}) we arrive at
\begin{widetext}
\begin{eqnarray}
&&\Im\Pi^{(T,ij)}_{(\rho,1)}(w,u)=g^{2}\int \frac{d^4l}{2(2\pi)^4}\,(n_{B}((l-w)\cdot u)+n_{B}(l\cdot u))(B^{[ll]}_{(T,ij)}\,(A_{\pi}^{[11]}(l,u)\,A_{\pi}^{[00]}(l-w,u)\nonumber\\[1.2 ex]
&&\qquad\qquad\qquad\qquad+2\,A_{\pi}^{[10]}(l,u)\,A_{\pi}^{[10]}(l-w,u)+A_{\pi}^{[00]}(l,u)\,A_{\pi}^{[11]}(l-w,u))+B^{[ww]}_{(T,ij)}\,(A_{\pi}^{[00]}(l,u)\,A_{\pi}^{[11]}(l-w,u))\nonumber\\[1.2 ex]
&&\qquad\qquad\qquad\qquad-(B^{[lw]}_{(T,ij)}+B^{[wl]}_{(T,ij)})\,(A_{\pi}^{[10]}(l,u)\,A_{\pi}^{[10]}(l-w,u)+A_{\pi}^{[00]}(l,u)\,A_{\pi}^{[11]}(l-w,u))\nonumber\\[1.2 ex]
&&\qquad\qquad\qquad\qquad+(B^{[ul]}_{(T,ij)}+B^{[lu]}_{(T,ij)})\,(A_{\pi}^{[20]}(l,u)\,A_{\pi}^{[10]}(l-w,u)+A_{\pi}^{[10]}(l,u)\,A_{\pi}^{[20]}(l-w,u)\nonumber\\[1.2 ex]
&&\qquad\qquad\qquad\qquad+A_{\pi}^{[12]}(l,u)\,A_{\pi}^{[00]}(l-w,u)+A_{\pi}^{[00]}(l,u)\,A_{\pi}^{[21]}(l-w,u))\nonumber\\[1.2 ex]
&&\qquad\qquad\qquad\qquad-(B^{[uw]}_{(T,ij)}+B^{[wu]}_{(T,ij)})\,(A_{\pi}^{[20]}(l,u)\,A_{\pi}^{[10]}(l-w,u)+A_{\pi}^{[00]}(l,u)\,A_{\pi}^{[12]}(l-w,u))\nonumber\\[1.2 ex]
&&\qquad\qquad\qquad\qquad+B^{[uu]}_{(T,ij)}\,(A_{\pi}^{[22]}(l,u)\,A_{\pi}^{[00]}(l-w,u)+2\,A_{\pi}^{[20]}(l,u)\,A_{\pi}^{[20]}(l-w,u)+A_{\pi}^{[00]}(l,u)\,A_{\pi}^{[22]}(l-w,u)))\nonumber
\end{eqnarray}
\begin{eqnarray}
&&\Im\Pi^{(T)}_{(\rho,2)}(w,u)=g^2\int \frac{d^4l}{2(2\pi)^4}(H^{[T,T]}\,\Im\,\Pi_{(T)}(l,u)+\sum_{ij=1}^{2}\,H^{[T,ij]}\,\Im\,\Pi_{(ij)}(l,u))\nonumber\\
&&\qquad\qquad\qquad\qquad\qquad \times A^{[00]}_{\pi}(l+w,u)(n_{B}((l+w)\cdot u)+n_{B}(l\cdot u))\nonumber\\
&&\Im\Pi^{(nm)}_{(\rho,2)}(w,u)=g^2\int \frac{d^4l}{2(2\pi)^4}(H^{[nm,T]}\,\Im\,\Pi_{(T)}(l,u)+\sum_{ij=1}^{2}\,H^{[nm,ij]}\,\Im\,\Pi_{(ij)}(l,u))\nonumber\\
&&\qquad\qquad\qquad\qquad\qquad \times A^{[00]}_{\pi}(l+w,u)(n_{B}((l+w)\cdot u)+n_{B}(l\cdot u))\,.
\end{eqnarray}
\end{widetext}
Finally the coefficients $B^{[ij]}_{(mn)}$ and $B^{[ij]}_{(T)}$ are to
be specified. We give the non-zero components only
\begin{eqnarray}
B^{[ll]}_{(mn)}&=&L^{\mu\nu}_{(mn)}(w,u)\,l_{\mu}\,l_{\nu}\nonumber\\
B^{[uu]}_{(mn)}&=&L^{\mu\nu}_{(mn)}(w,u)\,u_{\mu}\,u_{\nu}\nonumber\\
B^{[lu]}_{(mn)}&=&L^{\mu\nu}_{(mn)}(w,u)\,l_{\mu}\,u_{\nu}\nonumber\\
B^{[ul]}_{(mn)}&=&L^{\mu\nu}_{(mn)}(w,u)\,u_{\mu}\,l_{\nu}\nonumber\\[2ex]
B^{[ll]}_{(T)}&=&\frac{1}{2}T^{\mu\nu}(w,u)\,l_{\mu}\,l_{\nu}\nonumber\\
B^{[uu]}_{(T)}&=&\frac{1}{2}T^{\mu\nu}(w,u)\,u_{\mu}\,u_{\nu}\nonumber\\ 
B^{[lu]}_{(T)}&=&\frac{1}{2}T^{\mu\nu}(w,u)\,l_{\mu}\,u_{\nu}\nonumber\\
B^{[ul]}_{(T)}&=&\frac{1}{2}T^{\mu\nu}(w,u)\,u_{\mu}\,l_{\nu}\nonumber\\ 
B^{[uu]}_{(T)}&=&\frac{1}{2}T^{\mu\nu}(w,u)\,u_{\mu}\,u_{\nu}\\[2ex]
\label{B-coeff}
%
%
%
B^{[ll]}_{(T)}&=&\frac{1}{2}l^2-\frac{1}{2(w^2-\uq^2)}\left(\ql^2\right.
\nonumber\\
&& \quad\left.-2\,\ul\,\uq\,\ql+\ul^2\,w^2\right)\nonumber\\
B^{[ll]}_{(11)}&=&\frac{\ql^2}{w^2}\nonumber\\
B^{[ll]}_{(22)}&=&\frac{(\ul\,w^2-\uq\,\ql)^2}{w^2\,(w^2-\uq^2)} \nonumber
\end{eqnarray} 
\begin{eqnarray}
B^{[ll]}_{(12)}&=&B^{[ll]}_{(21)}=\frac{\ql\,(\uq\,\ql-\ul\,w^2)}
{w^2\,\sqrt{w^2-\uq^2}}\nonumber\\
%
%
 B^{[lu]}_{(22)}&=&B^{[ul]}_{(22)}=\ul-\frac{\uq\,\ql}{w^2}\nonumber\\
 B^{[uu]}_{(22)}&=&1-\frac{\uq^2}{w^2} \nonumber\\
 B^{[lu]}_{(12)}&=&B^{[ul]}_{(21)}=-\frac{\ql\sqrt{w^2-\uq^2}}{w^2}\nonumber\\
 B^{[qq]}_{(11)}&=&w^2\nonumber\\
 B^{[lu]}_{(21)}&=&B^{[ul]}_{(12)}=\frac{\uq\,(\uq\,\ql-\ul\,w^2)}{w^2\,\sqrt{w^2-\uq}}\nonumber\\[2ex]
%
%
 B^{[lu]}_{(11)}&=&B^{[ul]}_{(11)}=\frac{\uq\,\ql}{w^2}\\
 B^{[qu]}_{(11)}&=&B^{[uq]}_{(11)}=\uq\nonumber\\
 B^{[uu]}_{(12)}&=&B^{[uu]}_{(21)}=-\frac{\uq\sqrt{w^2-\uq^2}}{w^2}\nonumber\\
 B^{[lq]}_{(11)}&=&B^{[ql]}_{(11)}=\ql\nonumber\\[2ex]
%
 B^{[uu]}_{(11)}&=&\frac{\uq^2}{w^2}\nonumber\\
 B^{[lq]}_{(21)}&=&B^{[ql]}_{(12)}=\frac{\uq\,\ql-\ul\,w^2}{\sqrt{w^2-\uq^2}}\nonumber\\
%
 B^{[qu]}_{(12)}&=&B^{[uq]}_{(21)}=-\sqrt{w^2-\uq^2}\nonumber
\end{eqnarray}  


\end{document}